\documentclass[10pt,letterpaper]{article}
\usepackage[top=0.85in,left=2.75in,footskip=0.75in]{geometry}

\usepackage{amsmath,amssymb}

\usepackage{changepage}

\usepackage[utf8x]{inputenc}

\usepackage{textcomp,marvosym}

\usepackage{cite}

\usepackage{nameref,hyperref}

\usepackage[right]{lineno}

\usepackage{microtype}
\DisableLigatures[f]{encoding = *, family = * }

\usepackage[table]{xcolor}

\usepackage{array}

\usepackage[aboveskip=1pt,labelfont=bf,labelsep=period,justification=raggedright,singlelinecheck=off]{caption}

\makeatletter
\renewcommand{\@biblabel}[1]{\quad#1.}
\makeatother

\usepackage{lastpage,fancyhdr,graphicx}
\usepackage{epstopdf}
\fancyheadoffset[L]{2.25in}
\fancyfootoffset[L]{2.25in}
\usepackage{graphicx}
\usepackage{listings}
\usepackage{color}
\definecolor{dkgreen}{rgb}{0,0.6,0}
\definecolor{gray}{rgb}{0.5,0.5,0.5}
\definecolor{mauve}{rgb}{0.58,0,0.82}

\title{Introduction to Decentralization and Smart Contracts}
\author{Theodosis Mourouzis (theodosis@ciim.ac.cy)\textsuperscript{1,2} \\ Jayant Tandon (CIIM)\textsuperscript{1}}

\begin{document}
\maketitle
\textsuperscript{1}
Cyprus International Institute of Management, 21 Akademias Ave, Nicosia 2107, Cyprus\\

\textsuperscript{2}
UCL Centre for Blockchain Technologies, UCL Computer Science, Malet Place, London, WC1E 6BT, UK.\\\\
\textbf{Abstract}\\
The aim of this work is to study the use of decentralization and smart contracts on blockchain networks. We investigate the implementation and use of smart contracts on the platforms Bitcoin, Ethereum and Hyperledger Fabric. Additionally, we have researched consensus algorithms and their respective uses, mentioning both advantages and disadvantages where necessary. To conclude, there is an example contract that is meant to be a close to direct translation of a generic legal house rental contract to show how a legal contract can be translated.
\pagebreak

\section{Introduction/Summary}
	A Smart Contract is computerized transaction protocol intended to digitally facilitate, verify, or enforce the negotiation or performance of a contract. The term was coined by Nick Szabo who first proposed them. The aim of its design is to satisfy common contractual conditions, minimize exceptions that are both malicious and accidental as well as minimize the need for trusted intermediaries. While they may seem to be new, their conceptual design is based in basic contract law. Currently, the implementation of Smart Contracts is based on Blockchains. They are used more specifically in the sense of general-purpose computation that takes place on a blockchain or distributed ledger. 
	This article begins by introducing the concept of Decentralization and its uses along with a small discussion about its benefits and the consequences that accompany it. In addition, there is an in depth analysis of Consensus algorithms, which are essential in the development of a Blockchain network and the execution of Smart Contracts, introducing the most common ones and how they work. It further goes on to study their use and implementation in three Blockchain platforms: Bitcoin, Ethereum and Hyperledger Fabric. Each platform is briefly introduced and then a detailed analysis of how Smart Contracts are implemented and utilized on the platform.

\section{Decentralization}

	\begin{figure}[h!]
		\centering
		\includegraphics[width=0.7\linewidth]{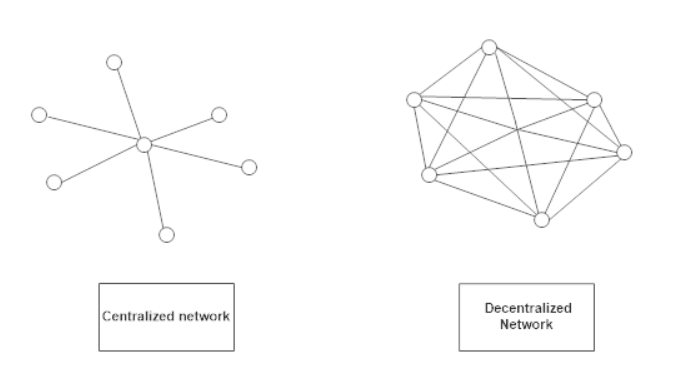}
		\caption[]{A depiction of Centralized and Decentralized networks}
		\label{fig:one}
	\end{figure}

	The proper definition of decentralization is: the transfer of authority and responsibility for public functions from the central government to subordinate or quasi-independent government organizations and/or the private sector. Blockchain technology is a very good example of this as it involves nodes that interact with each other directly instead of through a central node. This is also called Peer-to-Peer (P2P) networking. All peers/nodes have equal privileges and are equipotent participants of the network. They make a portion of their resources (e.g. processing power, network bandwidth, disk storage, etc.) to other network participants. The peers are both consumers and providers of the resources.
	
	\begin{figure}[h!]
		\centering
		\includegraphics[width=0.7\linewidth]{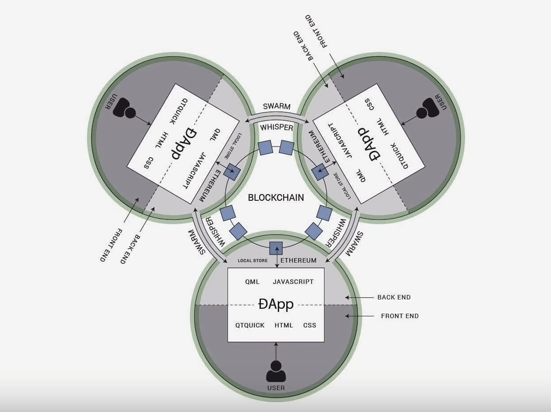}
		\caption[]{The structure of a Blockchain network}
		\label{fig:two}
	\end{figure}
	
	Blockchain, first implemented by Satoshi Nakamoto in 2008, is defined as a growing list of records (called blocks) that are linked in a chain using cryptography and are listed in an open, distributed ledger that is available to all nodes on the Blockchain network. Each block contains a cryptographic hash of the previous block, a timestamp and hashed transaction data (generally represented as a Merkle tree root hash). The design of each block is such that it is resistant to modification, which ensures its security and verifiability. It is such that although the blocks are not unalterable, to alter one block requires alteration of all subsequent blocks which requires network majority (i.e. for the majority of the nodes in the network to come to a consensus). Thus, the blockchains can be considered secure by design and exemplify a distributed computing system with high Byzantine Fault tolerance. The nodes in the network adhere to a set of rules (a protocol) which are defined and maintained by the consensus algorithm used by the network. These protocols mainly exist to enable inter-node communication and validation of new blocks.  
	
	\begin{figure}[h!]
		\centering
		\includegraphics[width=0.7\linewidth]{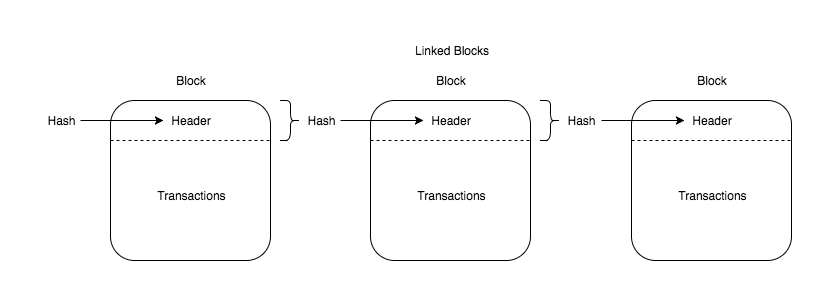}
		\caption[]{A depiction of a Block’s structure}
		\label{fig:three}
	\end{figure}
	
	Decentralization is a very impactful invention as it brings into play a way for people to communicate and exchange resources without the need for a central authority to govern the interactions. Such a system drastically reduces the impact an individual could have on it, as for every decision to be passed a consensus must be reached by the majority of the network. Thus, unless the majority of the network has malicious intent it is almost impossible for a malicious decision to be passed. The probability of this happening is dependent on the consensus algorithm of the network in question. Additionally, it drastically reduces the price of setting up a network as there is no need to set-up and maintain a high-cost mainframe along with back up systems. This is because each node acts as its own server and connects with other nodes to form the network. Since each node shares data even if one node goes offline the others still maintain the state of the network and the data is essentially backed-up in every single one of the nodes.\\
	Decentralization not only benefits the system but also the users. Since the nodes have equal authority and that the information of each node is not stored in a central location, even if one node of the network is hacked none of the other nodes are directly compromised. Additionally, due to the high interaction and integration of the nodes the possibly of a system hacker is very low. Decentralization also enables a greater level of openness and transparency in the network as all transactions are listed in the distributed ledger, so all nodes can check them and verify if the deals they have made with others have been carried out. This also means that the trust is dependent on each node and not on a central body. Thus, nodes are incentivized to act appropriately and in a trustworthy manner to ensure that they are still able to have other nodes trust them. Naturally some nodes still attempt to trick the system as can be exemplified by the case of Double-spending. Double-spending is a potential problem of the digital cash scheme where a single token can be used multiple times by duplicating or falsifying the digital file associated with the digit tokens. Double-spending creates similar issues to what counterfeit money does for fiat currencies. The main issue is the inflation it creates as new, fraudulent money is created and thus devalues the currency as well as damaging user trust in the currency and the circulation of the money. Many different approaches have been tried to prevent double spending. The most commonly used and accepted method is Proof of Work (PoW) which is further explained later in the text.\\
	Most of the implementations of decentralization use the systems theory approach. The systems theory is the study of systems. Systems are a cohesive conglomeration of interrelated and independent parts that are either natural or man-made. Each system is delimited by its spatial and temporal boundaries, surrounded and influenced by its environment, described by its structure and purpose or nature and expressed in its functioning. In the case of a decentralized network, the network is considered a system which is made up of multiple systems such that they work together to the benefit of the network.  In some sense each node could be considered its own system as each node works in tandem with other nodes to perform tasks defined by another system, such as the ordering service or the consensus algorithm.\\
	So, as can be seen decentralization is very beneficial to many, but it begs the question – what are the practical use cases of decentralization? First and foremost, would be the case of blockchain networks considering this paper is focused on it. Blockchain networks thrive on the fact that there is no need for a central authority to allow transactions to occur. This is because it reduces the cost of the transaction since there is no need for a commission to the structure that is executing the transaction. Decentralization is also of great use in political decision making as it would allow the citizens to be more active in the formulation and implementation of policies. This would allow the citizens to be more aware of policies that could affect them and make the policies more accepted and relevant to the citizens whom they affect. Another use of Decentralization would be one that is commonly seen: Privatization. Privatization involves leaving the provision of goods and services to the private sector and away from the public sector, meaning that the centralized government has less control and individuals (or groups of individuals) that are not part of the centralized governing body gain power over the provision of goods and services. Thus, achieving a rudimentary form of Decentralization. There are more examples and applications of Decentralization however those are details that will be glossed over in this article as only the basic understanding of it is required.
	
\section{Consensus Algorithms}
	Consensus algorithms are algorithms used to allow the nodes in a network to come to an agreement on a particular decision (i.e. achieve a consensus). Consensus algorithms are divided into two main levels: Crash Fault Tolerance (CFT) and Byzantine Fault Tolerance (BFT). CFT level algorithms ensure that the entire system can still function properly (and thus still reach a consensus) even if there are components which have failed. BFT level algorithms assume that every component has the potential to crash or be a source of malicious activity, thus it assumes a certain number of the nodes are malicious and ensures that the system will remain unaffected even if those nodes do turn out to crash/be a source of malicious activity. BFT algorithms are only applicable to distributed systems as the scenarios in which they can be applied must have decision making done by a group of individuals (or nodes) that are spread apart and do not have a central location which performs the decision making.  

\subsection{Order-Execute}	

	\begin{figure}[h!]
		\centering
		\includegraphics[width=0.7\linewidth]{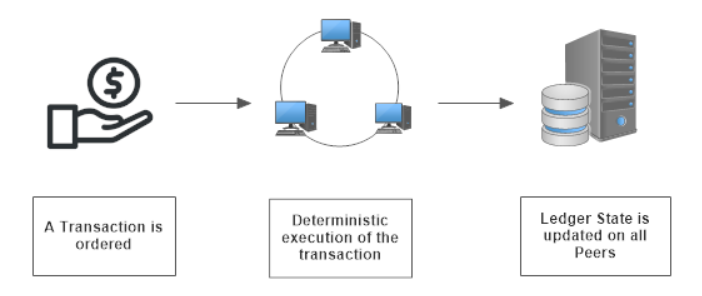}
		\caption[]{Simplistic description of the Order-Execute architecture}
		\label{fig:four}
	\end{figure}

	One implementation of consensus algorithms is the Order-execute architecture. This is a very common architecture used to maintain the network. The order-execute architecture is split into two phases: ordering phase, execution phase. In simple terms, the two phases occur as such: The network orders a transaction via a consensus protocol then the transaction is executed on every node simultaneously. The execution of transactions is sequential.
	Although the order-execute architecture is widely used, it has its own limitations as well. One such limitation is sequential execution, which means that only one transaction is executed on the network at any given moment. This leads to a limit on the effective throughput achievable by the network and thus can become a performance bottleneck. It also makes the network vulnerable to DoS (Denial of Service) attacks. This is easily accomplished as an attacker could very easily create a transaction which would loop infinitely and thus never finish executing. This would permanently freeze the network as there would be no more transactions being ordered since they are executed sequentially. \\
	Another limitation of the order-execute order is the fact that it needs to account for non-deterministic code. This is because transactions executed after consensus in active SMR (State Machine Replication) must be deterministic otherwise it will cause the network to “fork”. Forking is when the network produces two (or more) blocks at the same time, leading to multiple chains that could each be a valid and viable choice to follow. Thus, to resolve this issue there must be a way to decide which fork will be followed. In Bitcoin this is done by choosing between the forks and the one with the most votes will become the main fork while the other will be discarded. This generally means that there will be 51\% or more votes for the main fork as there are typically only two forks that are formed, although more are theoretically possible. This issue is generally resolved by using proprietary languages such as Solidity for Ethereum.
	Furthermore. there is the limitation of confidentiality of execution. This can be completely ignored in the case of a permisionless network, however in a permissioned network it is quite important. This is because public (permisionless) blockchains execute smart contracts on all nodes, however a permissioned network needs to have the ability to restrict access to smart contract logic, ledger state and/or transaction data. It is theoretically possible to achieve this using cryptographic techniques, however it is not applicable practically as it creates a large overhead. It is instead achieved by running the contract on a subset of the peers, to limit the execution of the smart contracts to a few trusted peers that can vouch for the validity of the smart contract. This is a variant of passive replication.  
	
	\subsection{Order-Execute-Validate}
	
	\begin{figure}[h!]
		\centering
		\includegraphics[width=0.7\linewidth]{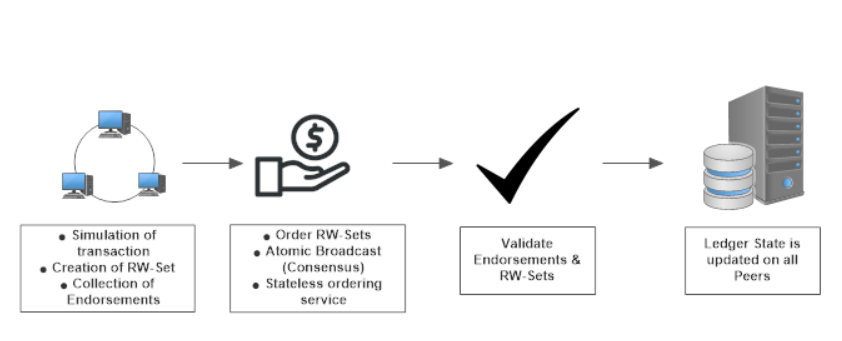}
		\caption[]{Simplistic description of the Execute-Order-Validate architecture}
		\label{fig:five}
	\end{figure}
	
	Another architecture is the “Execute-Order-Validate” architecture. It is used by HyperLedger Fabric and is meant to increase the modularity of the platform as well as bring other improvements (such as non-hard coded consensus) to the platform. Nodes in the network are separated into: Clients, Peers, and OSN (Ordering Service Nodes)/Orderers. The Clients are responsible for submitting transaction proposals for execution, to help orchestrate the Execution Phase, and to broadcast the transactions for ordering. Peers execute the transaction proposals and validate them. The Peers also maintain the ledger and state (a representation of the latest ledger state). A subset of the Peers, called Endorsing Peers/ Endorsers, have another job: to execute a transaction as specified by the Chaincode. There are three phases to the execute-order-validate architecture: Execution Phase, Ordering Phase, and Validation Phase.

	\begin{figure}[h!]
		\centering
		\includegraphics[width=0.7\linewidth]{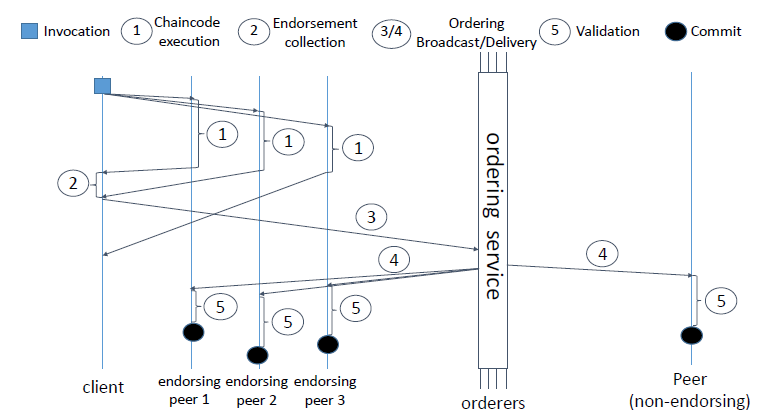}
		\caption[]{Execution of a Chaincode on HyperLedger Fabric}
		\label{fig:six}
	\end{figure}

	The Execution Phase involves the Clients signing and sending a transaction proposal to the Endorsers as specified by the Chaincode. The Chaincode contains: Identity of the Client; the payload in the form of an operation to execute; parameters; the identifier of the Chaincode; a nonce to be used by each Client; and the transaction identifier (which is derived from the client identifier and the nonce). \\
	The Endorsers simulate transaction proposal by executing operations on the Chaincode. The Chaincode runs in a Docker container – isolated from the main endorser process to ensure that any issues that may arise due to its execution will not affect the endorser’s other processes. The proposal runs against the local state of the Endorser and the Endorsers do not persist the results of the simulation to the ledger state. The state of the Blockchain is maintained by the PTM (Peer Transaction Manager). The state is stored in the form of a versioned key-value pair.\\
	For each unique key a tuple in the form (Key, Value, Version) is stored, with successive key-value pairs in monotonically increasing version numbers. The Version consists of the Block sequence number and the sequence number of the transaction within the Block. The PTM records what is accessed by the GetState as a tuple (Key, Version) in the Readset and what is accessed by the PutState as a tuple (Key, Value) in the Writeset for each entry. It ensures that the state created by a Chaincode is only directly accessible by the Chaincode that created it. The Chaincode is not required to maintain the local state of the program code, it only maintains the state that is accessed by the GetState, PutState and DelState operations. \\
	A Chaincode can invoke another Chaincode in order to access its state in the same channel as long as it is given the appropriate authority. Due to the simulation, each Endorser produces: a value WriteSet containing the state updates (modified Keys and their new values); a ReadSet which represents the version dependencies (the Keys read and their version numbers) of the simulation. After the simulation the Endorsers cryptographically sign a message called an Endorsement which contains: a WriteSet, a ReadSet, and other metadata. This Endorsement is returned to the Client in a Proposal Response. The Clients collect the Endorsements (on the basis that they are the same for each Endorser) until the requirements of the Endorsement Policy are met. Although this design helps simplify the architecture and is good enough for most Blockchain applications, it still has some issues. The main issue is that if there is high contention between nodes the Clients may be unable to satisfy the Endorsement Policy, thus the transaction will not be passed due to each node is considered untrustworthy on its own. \\
	The Ordering Phase is established on the total order of the submitted transaction per channel. In this phase the Endorsements are automatically broadcasted to establish consensus on transactions despite faulty orders. Multiple transactions are grouped into blocks and a hash-chained sequence of blocks. This helps improve the throughput of the broadcast protocol and is commonly used in Fault-Tolerant broadcasts. At a high level only two functions are supported: $broadcast(tx)$, $B <- deliver(s)$. These are invoked by a Peer and implicitly parameterized by a channel identifier. $broadcast(tx)$ is called by a Client to broadcast an arbitrary transaction “tx” containing the transaction payload and signature of the Client for dissemination. $B <- deliver(s)$ is called by a Client to retrieve the Block “B” with the non-negative sequence number “s”. The Ordering Service ensures the following safety properties of each channel:  Agreement, Hash Chain Integrity, No skipping, No creation. The agreement entails that for any two Blocks B and B’ delivered by sequence numbers $s$ and $s’$ respectively at correct Peers such that $s = s’$ the statement $B = B’$ is true. Hash Chain Integrity means that if a correct Peer delivers a Block B with sequence number s and another correct Peer delivers Block $B’ = ([tx1, tx2, …, txk], h’)$ with sequence number $s+1$ then it holds that $h’ = H(B)$ where $H(·)$ denotes the cryptographic hash function. \\
	This combined with the chaining of Blocks improves the efficiency of verifying the integrity of the Block sequence by Peers. No skipping states that when a correct Peer p delivers a Block with number $s > 0$ then for $(i = 0, 1, …, s-1)$ p has already delivered a Block with number $i$. Additionally, there is an “eventual” property that is also supported: Validity. Validity ensures that that if a correct Peer invokes $broadcast(tx)$ then every correct Peer eventually delivers a Block B that includes tx with some sequence number. There is also a built-in Gossip service for disseminating delivered Blocks to all Peers. This service is scalable and agnostic to the implementation of the Ordering Service, so it can work with both CFT and BFT, thus ensuring its modularity. The Ordering Service may occasionally perform access control checks to confirm whether a given Client is permitted to broadcast messages or receive blocks on the Channel. To further increase the modularity and to allow the consensus protocol to implement the ordering system the Ordering Service performs no validation or execution of transactions. \\
	The Validation Phase is split up into three steps: Endorsement Policy evaluation. Read-Write Conflict check, Ledger Update Phase. The Endorsement Policy evaluation is the validation of the Endorsement with respect to the Endorsement Policy. It happens in parallel for all Blocks and is the task of the VSCC (Validation System Chaincode). The VSCC is a static library that is part of the Blockchain’s configuration. If the Endorsement Policy is not satisfied the transaction is marked as invalid and is not passed. The Read-Write Conflict check is a comparison of the keys in the ReadSet field with those in the current state of the ledger (as locally stored by the Peer). It checks whether or not the keys are identical, if they are not identical then the transaction is marked as invalid and is not passed. This is done for all transactions in a Block sequentially. The Ledger Update Phase involves the new Block being appended to the locally stored ledger and then an update to the Blockchain’s state using the results of the previous steps. The state updates are applied by writing all Key-Value pairs in the WriteSet to the local state. This helps facilitate a reconstruction of the state later. The ledger contains all transactions, even invalid ones. This helps track the transactions and can be used to identify DoS attacks and put in preventive measures such as blacklisting the nodes that proposed these transactions.\\
	There are a few other architectures that exist, however each comes with its own limitations and advantages. Most of these architectures (for permissioned networks) implement the Fixed Trust model, where they use asynchronous BFT replication protocols to establish consensus. These protocols generally rely on a security assumption that among $n > 3f$ peers, where $n$ is the total number of nodes on the network, $f$ peers are tolerated to misbehave and show byzantine faults. However, this makes the system less flexible as the trust at the application level is fixed to trust at the protocol level.\\
	Another limitation of existing architectures is the hard-coded consensus. This is a very common practice in both permissioned and public blockchain networks as it is easier to implement and makes a standard for the network that everyone can use, thus there is little confusion about it. However, it comes with its obvious set of limitations for the network as well, namely that there is simply no “one size fits all” solution that can be applied to every scenario, so the consensus limits the network in ways that are very specific to the situation and not just a general limitation for the network.\\
	Additionally, most consensus algorithms that use CFT or BFT are bandwidth bound. This means that the throughput of the ordering service is capped by the network capacity of its nodes, thus consensus cannot be scaled by simply adding more nodes and will instead cause the throughput to decrease.\\
	There are many different protocols that are used by blockchain networks. One protocol is Proof of Work (PoW). The most popular implementation of this protocol is Bitcoin. In this protocol the nodes are called ‘miners’ as they compete to be the first to solve cryptographic puzzles to create a new block which is added to the network. Whenever a miner succeeds in solving a puzzle it is rewarded with the network’s currency. This is known as a block reward. Each new block must follow a set of consensus rules, those that do not are rejected by the network and thus are discarded. The way that miners solve these puzzles is by taking data from the block header as input and repeatedly running it through a cryptographic hash function. The hashes are designed in a way that ensures they are hard to find but easy to verify.\\
	The miners hash slight variations of the input data by including a nonce each time the data is run. The correct hash is found when a defined number of leading zero bits is found. Generally, the difficulty of the puzzles is designed to increase as more are solved, however if consecutive puzzles are solved too quickly or too slowly there will be an adjustment to the network difficulty. This is the measure of how difficult it is to find a hash below a target, and it is dynamically adjusted to ensure that the rate ate which new blocks are generated remains constant. The biggest issue with PoW is that the difficulty of creating new blocks increases infinitely, thus it requires increasingly greater computational power. This makes it very improbable for miners to generate new blocks, making it harder for them to earn any reward. Thus, it becomes less profitable for the miners. Thus, mining can only occur profitably and reasonably with large numbers of ASICs that are dedicated to mining. This means that decentralization is reduced as the main miners that remain are only the ones that can afford to do so, which is much fewer in number.\\
	Another protocol is Proof of Stake (PoS). The most popular implementation of this protocol is Ethereum. In this protocol a few nodes are chosen to generate a new block for the network. The chosen nodes are called Validators and are chosen based on their economic stake on the network. The economic stake can consist of: The relative value of the coins held in the validator’s wallet, the coin age of the tokens in the validator’s wallet. The relative value is found by dividing the total value of the coins in the wallet by the total value of coins in the network. The higher the relative value, the more likely the node is to be chosen as a validator. Coin age is defined as the coin amount (in the wallet) multiplied by the number of days it has been in the wallet. A higher coin age gives a greater chance to be chosen as a validator.\\
	A validator generates a new block by sending a special transaction that also locks up their deposit. The deposit servers as a collateral to ensure that the validator has no incentive to misbehave (i.e. do anything that would negatively impact the network) during the block generation process. If the validator is caught misbehaving, e.g. by trying to cheat the system by validating fraudulent transactions, the deposit will be slashed. The validators that correctly validate blocks are returned their deposit and an additional transaction fee.\\
	The main problem with PoS is the Nothing at Stake problem. This is in the case of an event where two blocks are produced at the same time (known as ‘forking’), thus resulting in two competing blockchains. The validators are incentivized to form more blocks on top of their blockchain to ensure it will eventually win – i.e. become the main fork. The problem with this is that, if the validators are presumed to be economically rational, then no chain will win as the same stake can be applied to each chain. Thus, it would create a risk-free method for validators to increase their rewards. This problem can be resolved by ‘slashing’ – penalizing validators who form blocks on multiple chains simultaneously. PoS is considered better than PoW as PoS has a considerably smaller energy consumption.\\
	Additionally, there exists the Delegated Proof of Stake (DPoS) protocol. One platform that uses this is BitShares. This protocol is quite different from others and instead is similar to the method in which elections occur in the US. The coin holders (nodes) vote for delegates who are responsible for validating transactions and maintaining the blockchain, including the ability to propose changes to the network. Once the changes have been submitted, it is up to the stakeholders to decide if they will be implemented. This is like electing Senators for the Senate. The stakeholders also elect witnesses, who generate new blocks and are rewarded for doing so. Stakeholders can only vote once per witness; however, they can vote for any number of witnesses as long as at least 50\% of the stakeholders believe sufficient decentralization has been achieved. The voting for witnesses is continuous, so witnesses are incentivized to perform to the best of their ability to ensure they are not replaced. Typically, there will also be a fixed time at which a chosen group of witnesses will be replaced. To aid this voting process there exists a reputation scoring system which assesses the quality of the witnesses. If a witness fails to generate a block there will be a negative impact on their reputation score. DPoS has both its advantages and disadvantages over other protocols, such as its far lower energy consumption and is better promotes decentralization than PoW, however DPoS can never achieve sufficient decentralization.\\
	There is another protocol, called Proof of Importance (PoI), that is used. This was introduced by NEM and is also used on it. PoI is the mechanism through which the network decides which nodes are to be chosen for the task of adding blocks to the blockchain. This process is called “harvesting” in PoI. The nodes that harvest blocks are able to collect the transaction fees in exchange for the resources, computing power, they use to perform the task. To choose the nodes that will harvest, three factors are considered. These factors take into account the total support a node has given to the network as a whole. The first factor is the amount of coins a node has vested in the network, with a minimum number of coins that have been in the wallet for a set number of days being required to be qualified to harvest and every coin after the minimum increases the chance of being selected. The second factor is the number of transaction partners a node has, i.e. how many other nodes has it made transactions with. The more transaction partners a node has the higher the likelihood of it being selected to harvest is. To avoid exploitation by nodes having transactions back and forth between themselves only the net transfers over time are accounted for. The third factor is the number and size of transactions over a set number of days. Each transaction above a minimum size contributes to this. The larger and the more frequent transactions a node has the higher its chance of being chosen to harvest.
	
\section{Smart contracts development simplified}
\subsection{Bitcoin}
	Bitcoin is the most well-known cryptocurrency thanks to the massive increase in value it had just recently, where it reached an exchange rate of \$19,383.06 per 1 BTC. It was the first mainstream cryptocurrency and the pioneer of PoW. It was originally created as only a digital currency platform and thus compared to other platforms smart contracts are not as programmable and extensible. The primary responsibility of the Bitcoin network is to maintain its distributed ledger. 
	The language in which smart contracts are executed is called Bitcoin Script. Bitcoin Script is a low-level language, thus it is quite difficult and requires much more time to write a contract using it. Instead there are high-level languages such as Ivy-Lang and Balzac that are used to write the contracts which then compile into Bitcoin Script. In this article we will only look at Ivy-Lang and Bitcoin Script, the official documentation for Balzac is available online with all the details required.\\
	As stated previously, Bitcoin Script is a low-level proprietary language that is used to write the programs of a smart contract. Bitcoin Script consists of a series of opcodes that are executed by a stack based Virtual Machine (VM). It supports both cryptographic and hash functions, as well as conditionals. However, it is not Turing-complete since it does not support looping or recursion. For security, confidentiality and consistency, Bitcoin Script does not allow scripts to inspect other inputs or outputs, so the contract cannot control the flow of value nor maintain any persistent state. Bitcoin Script is mostly used to create multisignature addresses, time-locked transactions, payment channels and cross-chain atomic trades.\\
	Ivy-Lang is a language that is meant to allow easier creation of smart contracts. It has a fairly regular syntax: Each contract must be passed some arguments, each of which has its own type. The contract must be parameterized by a cryptographic publicKey to create an address. There must always be a Value passed in the argument, as it represents the Bitcoins that are to be protected and transferred once the contract’s requirements are fulfilled. Each contract has multiple clauses of which at least one must be invoked, passed clause arguments and satisfy each of its conditions to unlock the contract. Each clause must also end with a statement such as unlock Value to unlock the contract. The supported variable types in Ivy-Lang are:
	\begin{itemize}
		\item Bytes: string of bytes (typically in hexadecimal)
		\item PublicKey: ECDSA public key; Signature: ECDSA signature by some private key on the hash of the transaction; Time: either a block height or a timestamp
		\item Duration: either a number of blocks or a multiple of 512 seconds
		\item Boolean: true or false
		\item Number: integer between -2147483647 and 2147483647, inclusive
		\item Value: some amount of Bitcoins - parameters of Value type represent Bitcoins that are locked up in a transaction
		\item HashableType: any type that can be passed to a hash function: Bytes, PublicKey, Sha256(T), Sha1(T) and Ripmend160(T)
		\item Sha256(T: HashableType): result of taking a SHA-256 hash of a value of the hashable type T
		\item Sha1(T: HashableType): result of taking a SHA-1 hash of a value of the hashable type T
		\item Ripemd160(T: HashableType): result of taking a Ripmend-160 hash of a value of the hashable type T.
	\end{itemize}
	The functions and operators that are supported are listed as follows:
	\begin{itemize}
		\item $checkSig(publicKey: PublicKey, sig: Signature) -> Boolean$: checks that the signature sig is a valid signature on the spending transaction by the private key corresponding to publicKey
		\item $checkMultiSig(pulbicKeys: [PublicKey], sig: [Signature]) -> Boolean$: checks that at each signature in sig is a valid signature on the spending transaction by the private key corresponding to one of the publicKeys. The signatures must be provided in the same order as their respective public keys.
		\item $after(time: Time) -> Boolean$: checks that the current block time (or block height) is after time – uses the transaction’s nLockTime field and the CHECKLOCKTIMEVERIFY instruction
		\item $older(duration: Duration) -> Boolean$: checks that the contract being spent has been on the blockchain for at least duration (w). This uses the input’s sequence number and the CHECKSEQUENCEVERIFY instruction
		\item $sha256(preimage: (T: HashableType)) -> Sha256(T)$: computes the SHA-256 hash of preimage
		\item $sha1(preimage: (T: HashableType)) -> Sha1(T)$: computes the SHA-1 hash of preimage
		\item $ripemd160(preimage: (T: HashableType)) -> Ripemd(T)$: computes the RIPEMD-160 hash of preimage
		\item $bytes(item: T) -> Bytes$: coerces item to a bytestring. Has no effect on the compiled output or on script execution (since the Bitcoin Script VM treats every item as a bytestring), instead it only affects typechecking. Cannot be called on an item of type Value or Boolean)
		\item $size(bytestring: Bytes) -> Number$: gets the length of bytestring
		\item $==; =! $: checks the equality of any 2 values of the same type. Cannot be called on Booleans due to limitations of Bitcoin Script
		There are also some special contracts that are used to allow specific functions on the Bitcoin network. One such class of methods is called Colored Coins. This term describes a class of methods for representing and managing real world assets on top of the Bitcoin Blockchain. It does so by purposely forking the network to create a new blockchain entirely.
	\end{itemize} 

\subsection{Ethereum}
	Ethereum is another digital currency platform. It uses Solidity as the language in which smart contracts are written. Solidity is a Turing-complete language, i.e. it can compute anything computable as long as enough resources (computational power) are provided. Consensus in the network is reached through PoS, at the ledger level. Thus, it is shared among all participants in the ledger. The transactions are enforced and validated by the validators. They are paid using “gas” to complete their task. Gas is the unit by which the degree of difficulty of computational efforts is measured, i.e. the cost a validator has to execute a contract. Each contract has a . Gas transferred in ether.\\
	Ethereum is much more flexible than most other platforms. This is advantageous to the network as it allows better implementation of contracts and more types of contracts to be executed, however in some ways it is too flexible as it leads to unpredictability and lower safety of the platform. This is especially so because it allows for so many different approaches to the same goal, the tricks in the details can catch out even the most seasoned user.
	\begin{itemize}
		\item To reduce the confusion and make it easier and simpler for users to create reliable, safe contracts as well as deploy them some standards were developed. The most common standard is ERC20, which was first introduced in 2015. It is a basic standard and has its own issues, however it performs it functions well enough and is simple enough for the majority of users to use. The functions that ERC20 enables are: totalSupply – gets total supply of tokens
		\item $balanceOf(address\textunderscore owner) constant returns (unit256 balance)$ – retrieves total balance of another account associated with “\textunderscore owner”
		\item $transfer(address\textunderscore to, unit256\textunderscore value) returns (bool success)$ – sends a specific amount of tokens \textunderscore value to an address
		\item $transferFrom(address\textunderscore from, address\textunderscore to, unit256\textunderscore value) returns (bool success)$ – sends a specific amount of tokens \textunderscore value from one (token contract) address to another
		\item $approve(address\textunderscore spender, unit256\textunderscore value) returns (bool success)$ – allowing an account to withdraw tokens from another account with a predefined upper limit \textunderscore value
		\item $allowance(address*\textunderscore owner*, address*\textunderscore spender*) constant returns (unit256 remaining)$ – returns the number of remaining tokens, within the preset defined by the upper limit of the amount that the \textunderscore spender can spend from the account of \textunderscore owner
	\end{itemize}

	The main issue with ERC20 is that it assumes two ways of transaction: \\
	1.	Use transfer function to transfer tokens to another account\\
	2.	Deposit tokens using a combination of approve and transfer\\
	However, it is very easy to accidentally deposit tokens using transfer. This will succeed, but it will not be recognized by the recipient contract. This means that the exchange token balance of the account that transferred the tokens will not be credited. Thus, the tokens will be permanently lost in the case of no emergency token extraction function is in place.\\
	To solve this issue, Reddit user u/Dexaran created the ERC223 standard, which is focused on security primarily. This standard allows token transactions to behave as ether transactions. It solves the issue of ERC20 by making the transfer function throw an error on invalid transactions, thus cancelling the transaction. This is done in two steps: the receiving address is checked to see whether it contains data (i.e. it is a contract), if it does contain data then it is assumed that there is a tokenFallback function. Of course, if no tokenFallback function exists then the contract’s fallback function is called and tokens can still be lost, however it is very unlikely for this to occur.\\
	There are many other standards as well. The main ones that will be mentioned with some details are: ERC20, ERC223 and ERC777. Two of these have already been mentioned with some details. ERC777 is another standard, one which standardized Mint \& Burn functionality of tokens. ERC777 creates a compromise between security and acceptability. It is based off ERC820 and focuses primarily on adoption by offering a large range of transaction handling mechanisms. It adds three new functions:
	\begin{itemize}
		\item $Send$ – replaces transfer
		\item $authoriseOperator$ – replaces approve
		\item $tokensRecieved$ – replaces tokenFallback
	\end{itemize}
	ERC777 uses a new way of recognizing the contract interface which assumes that there is a central registry of contracts on Ethereum’s network. This is based off ERC820. This central registry can be invoked by anyone to check if a specified address supports a specified set of instructions. Even though ERC777 brings in quite a few new things, it can still create a token using ERC20’s default functions with ERC777 functions without any overrides. This is the biggest reason for the ease at which ERC777 can be adopted.\\
	ERC777 is still not perfect and has its fair share of problems as well. One issue is that the authorizeOperator function is depreciated, worsens network bandwidth and requires more gas than its alternatives. This is because authorizeOperator acts as a transaction and then authorize withdrawal is another transaction, thus two transactions have to be executed for the contract to be used. Additionally, the flag that prevents stuck tokens, by performing checks about the ITokenRecipient interface and checking if the address is whitelisted, is optional while it should be compulsory to truly make the standard secure. Furthermore, although the central registry of the network brings about many advantages, it can also act as the central point of failure. This would greatly harm the entire network and thus it should be ensured that it is secure.\\
	Many other standards exist; however, they will not be discussed in detail here but some will be listed and briefly described as follows:
	\begin{itemize}
		\item ERC827 – extension of ERC20 allowing for calls inside transactions and approvals
		\item ERC664 – allows for modularity to be introduced over time and make tokens upgradable by abstracting user balances from token business logic
		\item ERC721 – standard interface for deeds (non-fungible tokens)
		\item ERC677 – Allows tokens to be transferred to contracts and have the contract trigger logic to decide on how to respond to the incoming tokens within a single transaction
		\item ERC820 – Defines a universal registry smart contract where any address can register which interface it implements and which smart contract is responsible for its implementation. Keeps compatibility with EIP-165
	\end{itemize}
\subsection{Hyperledger Fabric}
	Hyperledger Fabric, which is hosted by the Linux Foundation, is a blockchain network that is unique. First, it is aimed towards enterprise use and not individual, thus it is permissioned unlike the majority of blockchains. Secondly, it has no native cryptocurrency but instead gives the users the ability to create their own tokens that can be used on the platform. Thirdly, it is completely modular in its architecture, including its consensus. The modularity of its architecture allows it to be more scalable, fast and efficient compared to other public blockchains. \\
	In Hyperledger fabric Smart Contracts are called “Chaincode”, thus will be referred as such in this section. Chaincodes have 4 main functions:
	\begin{itemize}
		\item $PutState$ – Creates a new asset or update existing
		\item $GetState$ – Retrieves an asset
		\item $GetHistoryForKey$ – Retrieves history of changes
		\item $DelState$ – Deletes asset 
	\end{itemize}
	These are the main functions that are used for the smart contracts, however more exist. \\
	Hyperledger Fabric uses a state database which stores keys and their values, however the deletion of a key and value pair will have no effect on the blocks in the blockchain. It is instead treated as a transaction; thus, the history of a key can still be retrieved even if it is deleted. \\
	Thanks to the use of its execute-order-validate architecture, many Chaincodes can be run at the same time and be deployed dynamically (by anyone). However, all Chaincodes should assume that application code is not to be trusted – to reduce the chance of Byzantine faults. \\
	The execution of Chaincode occurs within a contained environment which is loosely coupled with the rest of the peers in the network. This allows the use of plugins to enable many languages, which are not proprietary, as the peers are not concerned about which language is used for implementation of the Chaincode. Currently Go, Java and Node are supported. Chaincode and peers communicate using gRPC (an open source remote procedure call system) messages. However, special Chaincodes known as System Chaincode are an exception. This is because they run directly in the peer process. Thus, they allow implementation of specific functions required by Hyperledger Fabric. They can also be used in the case that regular Chaincodes are too restrictive due to their isolation.\\
	System Chaincode, alongside Channel Configuration, are used to customize Hyperledger Fabric. Configuration of a channel is maintained in special Configuration Blocks. Each contains the full channel configuration and no other transactions. Each blockchain will begin with a configuration block - named the Genesis Block. The Genesis Block is used to bootstrap the channel. It includes:
	\begin{itemize}
		\item Definitions of the MSPs of the participating nodes
		\item Network addresses of the OSNs
		\item Shared configuration for the consensus implementation and the ordering service
		\item Rules governing access to the ordering service
		\item Rules governing how each part of the channel configuration may be modified
	\end{itemize}
	A channel can be configured through the use of a Channel Configuration Update Transaction. It contains: a representation of the changes that are to be made; a set of signatures that are used by the OSNs to validate and authenticate the transactions using the current configuration. The peers which receive this block check if the update is authorized based on the current configuration and then update the configuration if it is valid.\\
	General user-level Chaincodes are also known as Application Chaincode. They are deployed with a reference to ESCC (Endorsement System Chaincode) and to a VSCC (Validation System Chaincode). This is done so that the output of the ESCC is can be used as the input to the VSCC. The ESCC takes a proposal, and proposal simulation results as an input. If the results are satisfactory then the ESCC produces a response in the form of an endorsement. For the default ESCC an endorsement is simply a signature using the peer’s local signing identity. The VSCC takes a transaction as input and simply outputs a flag that indicates whether the transaction is valid or not. For the default VSCC the validation is by collecting the endorsements (as output by the ESCC) and evaluates them against the endorsement policy specified for the Chaincode.
	\pagebreak
\section{Real Life Example Coded}

\begin{lstlisting}
pragma solidity ^0.5.0; //indicates the compiler version to be used
import "github.com/oraclize/ethereum-api/oraclizeAPI.sol";

contract HouseRental is usingOraclize{ 
//Initiate the contract and inherit from the Oracle service Oraclize

 struct RentsPaid {
  uint month;
  uint amount;
 }
 enum Status {Created, Active, Terminated}

 uint daysInMonth;
 Status public status;
 address payable public landlord;
 address payable public tenant;
 string public house; 
  //Identifies the house that is to be rented, can be done using the address
  // of the house
 RentsPaid[] public rentsPaid; 
  //array of uint to store the rents that are paid
 uint public timeCreated; 
  //indicates the time on the block that the contract has been deployed
 uint public termLength;
  //The number of months for which the house will be on rent
 uint public rent = 1 ether;
  //Declares the rent that is to be paid, "1 ether" is just an example
 uint public securityDeposit = 1 ether;
 uint public lateFee;
 bool termsBreached;

 modifier onlyTenant() {
  assert(msg.sender == tenant);
  _;
 }

 modifier onlyLandlord() {
  assert(msg.sender == landlord);
  _;
 }

 modifier policyBreached() {
  assert(!termsBreached);
  _;
 }

 event rentPaid();
 event contractActive();
 event contractTerminated();
 event termBreached();
 event LogNewOraclizeQuery(string description);

 /* Getter functions just to ensure we can 
 "get" the value of any variable when needed
 */

 function getStatus() view public returns(Status) {
  return status;
 }

 function getLandlord() view public returns(address) {
  return landlord;
 }

 function getTenant() view public returns(address) {
  return tenant;
 }

 function getHouse() view public returns(string memory) {
  return house;
 }

 function getTimeCreated() view public returns(uint) {
  return timeCreated;
 }

 function getTermLength() view public returns(uint) {
  return termLength;
 }

 function getRent() view public returns(uint) {
  return rent;
 }

 function getSecurityDeposit() view public returns(uint) {
  return securityDeposit;
 }

 /* Cannot implement this function at the moment as it is unsupported:

 function getRentsPaid() view public returns(RentsPaid[] memory) {
  return rentsPaid;
 }

 */

 constructor (uint _rent, string memory _house, uint _termLength) public { 
 //Initiates the constructor function
  landlord = msg.sender; 
  //indicates the landlord's address, thus it should be the landlord who 
  //first deploys the contract
  rent = _rent;
  house = _house;
  termLength = _termLength;
  timeCreated = block.timestamp;
  termsBreached = false;
 }

 function beginLease() public payable policyBreached{
  require(msg.sender != landlord && status == Status.Created && msg.value ==
             securityDeposit);

  tenant = msg.sender;
  landlord.transfer(msg.value);
  status = Status.Active;
  emit contractActive();
 }
 function CheckTerms() public payable returns(bytes32){
  if(oraclize_getPrice("URL") > address(this).balance) {
   emit LogNewOraclizeQuery("Oraclize query was NOT sent, please add some ETH to cover for the query fee");
  } else {
   emit LogNewOraclizeQuery("Oraclize query was sent, standing by for the answer..");
   return oraclize_query(60*3600, "URL", "INSERT TERMS HERE, please respond in the format: true / false");
  }
 }

 function __callback(bytes32 myid, bool result) public {
  if (msg.sender != oraclize_cbAddress()) revert();
  if(CheckTerms() == "true"){
   result = true;
  } else {
   result = false;
  }
  termsBreached = result;
 }

 function payRent() public payable onlyTenant {
  require(status == Status.Active);
  require(msg.value == rent + lateFee);

  landlord.transfer(msg.value);
  rentsPaid.push(RentsPaid({
  month : rentsPaid.length + 1,
  amount : msg.value
  }));

 }

 function TerminateContract() public payable onlyLandlord {
  if(!termsBreached) {
   tenant.transfer(securityDeposit);
  }
  emit contractTerminated();
  selfdestruct(landlord);
 }

 function tokenFallback() public {
  landlord.transfer(address(this).balance);
 }
 function () external {
  tokenFallback();
 }
}
\end{lstlisting}
\pagebreak

\textbf{\textit{References}}
\begin{enumerate}
	\item Michiel Mulders. Comparison of Smart Contract Platforms.\\ https://hackernoon.com/comparison-of-smart-contract-platforms-2796e34673b7	
	\item Tetiana Boichenko. Top 3 Platforms for Successful Smart Contract Development\\
	https://www.n-ix.com/top-3-platforms-successful-smart-contract-development/
	\item Alyssa Hertig. How Do Ethereum Smart Contracts Work?\\
	https://www.coindesk.com/information/ethereum-smart-contracts-work/
	\item https://hyperledger-fabric.readthedocs.io/en/release-1.1/chaincode.html
	\item https://github.com/hyperledger/fabric
	\item https://github.com/ibm-watson-iot/blockchain-samples/blob/master/docs/iotcp/\\HyperledgerContractsIntroBestPracticesPatterns.md
	\item https://blockgeeks.com/guides/solidity/
	\item Elli Androulaki, Artem Barger, Vita Bortnikov, Christian Cachin, Konstantinos Christidis, Angelo De Caro, David Enyeart, Christopher Ferris, Gennady Laventman, Yacov Manevich, Srinivasan Muralidharan, Chet Murthy, Binh Nguyen, Manish Sethi, Gari Singh, Keith Smith, Alessandro Sorniotti, Chrysoula Stathakopoulou, Marko Vukolić, Sharon Weed Cocco and Jason Yellick. Hyperledger Fabric: A Distributed Operating System for Permissioned Blockchains\\
	https://arxiv.org/pdf/1801.10228v2
	\item http://thesecretlivesofdata.com/raft/
	\item https://raft.github.io/
	
\end{enumerate}

\end{document}